# When does the Richardson-Lucy deconvolution converge ?

Hongguang Bi[1,2], and Gerhard Börner[1]

[1] Max-Planck-Institut für Astrophysik, Karl-Schwarzschild Str. 1, D85748 Garching, Germany
[2] Department of Physics, University of Arizona, Tucson, AZ 85721, USA

**Abstract.** We propose a simulation-based bootstrap method to access global significance levels of deconvolution models in the Richardson-Lucy and other iterative restoration algorithms that converge locally. These significance levels allow one to check at each iterative step how good the model is and when iterations can be stopped. Adding more iterations in the deconvolution improves the fitting but is very slow at later time; while too much entropy or smoothness will be lost in the models. A good deconvolution model should firstly have a significance level as high as possible ($\geq 20\%$), and secondly, be as smooth as possible. We have used two examples to illustrate how such models can be derived in practice.

We point out that maximizing the sum of the likelihood of fitting and *a priori* entropy does not guarantee an acceptable significance level for the resulting model. If one's *a priori* knowledge is too poor, the model may not be able to fit the data at a reasonable significance level. Instead, a maximum-entropy-like iterative restoration algorithm can be performed later by acquiring *a priori* knowledge from the Richardson-Lucy restoration. However, this is necessary only when it does increase the levels significantly.

## 1. Introduction

Image restoration, spectral deconvolution and related inverse problems are found in many astronomical data processings such as Hubble Space-Telescope observations, X-ray data from the Einstein Observatory and ROSAT, and many radio observations etc.. A large number of deconvolution methods proposed so far are based on the idea called regularization which searches for best fitted models subject to constraints like maximum entropy, flatness or smoothness. A remarkable exception to the regularizations is the Richardson-Lucy iterative method, proposed in optics by Richardson (1972) and independently in astronomy by Lucy (1974). This method starts usually from an initial model of constant density distribution that apparently has maximum entropy as well as being perfectly smooth, then it modifies the estimates step by step by collecting information from the observational data until a reasonable fitting is reached. Roughly speaking, it uses a procedure just opposite to regularization.

Besides its nice mathematical properties, the main motivation for the Richardson-Lucy algorithm in recent years is due to its power in restoring HST images and its potential use in further satellite-based digital data. As an important application, the Richardson-Lucy method gives a so-called Object-Feasibility filtering that removes pixel-to-pixel Poisson noise but does not degrade the original detector resolution. This property is very useful in smoothing optical CCD images and X-ray images. A comprehensive study of the Richardson-Lucy algorithm and other related iterative deconvolution algorithms was given by Meinel (1986). Despite many practical successes, these methods have yet left a number of questions to be answered, among which the most important one may be to determine at which step the iteration should stop when the signal estimate can be said to be *good* or *convergent*. Since the likelihood of model fitting is improved after each step while an infinite number of iterations is neither possible nor useful (see below), this question must be studied carefully especially when *a priori* knowledge about the signal is not available or when the signal-to-noise ratio is poor.

The present paper will focus on these questions. First, we will propose a simulation-based bootstrap method to calculate the probability significance levels of estimated signal models in the Richardson-Lucy algorithm, which, though easily being an extension of discussions in Lucy (1974) and Meinel (1986), can be achieved in practice only with a large number of computer simulations. The goodness-of-fit test given in this article is a global measure of the algorithm that converges locally. Second, when the significant levels are available, we can define the conver-gent solution(s) quantitatively and measure how it, as well as entropy or smoothness, changes in the iteration. Thus one can easily decide how to incorporate entropy information into the deconvolution. In section 2, we first discuss general aspects of both regularization methods and the Richardson-Lucy method. Then we propose a bootstrap procedure to calculate significance levels and discuss some properties that were not addressed clearly before in the algorithm. We study a simple artificial model and a real ROSAT 1-D spectrum in section 3 for application illustrations. Application to 2-D HST or ROSAT images will be published elsewhere. We summarize the results in section 4.

## 2. Regularizations and Iterative Iterations

Let the observed counts be $c_i(i = 1, ..., N)$, the underlying signal $u_j(j = 1, ..., M)$ and the noise $n_i$. We have

$$c_i = \sum_{j=1}^{M} p_{ij} u_j + n_i, \quad i = 1, ..., N \quad (1)$$

here $p_{ij}$ is a point spread function (PSF). The observation can either be a 1-D spectrum or a 2-D image. Oversampling cases, say $M \geq N$, are allowed because sometimes the PSF can be measured in finer grids.

There are two types of noise (Lucy 1992). Type I noise is not associated with the true signal but comes from backgrounds such as the read-out noise of a detector or the sky background. This kind of noise appears like a smooth background and can be removed somehow before other data processing (e.g., several tools for this perpose have been implemented in astronomical softwares IRAF and MIDAS). In contrast, type II noise is associated with the signal recording process. A typical example is the pixel-to-pixel noise due to photon Poisson statistics in CCD detectors. Throughout this paper we assume that type I noise has been corrected in the data, only the Poisson noise will be considered.

### 2.1. Regularization Method

If there is a model estimate, $\tilde{u}_i$, an object-feasibility filtering $\tilde{c}_i$ is derived through the equation

$$\tilde{c}_i = \sum_j \tilde{u}_j p_{ij}, \quad (2)$$

and $\chi^2$ measures the goodness-of-fit :

$$\chi^2 = \sum_i \frac{(c_i - \tilde{c}_i)^2}{\tilde{c}_i}. \quad (3)$$

Apparently, the smaller $\chi^2$ is, the better the model agrees with the data. The best model can be solved by minimizing $\chi^2$. However, this model and models that are very close to it are usually very irregular even among the nearest pixels, which a real physical model should not be. Regularization methods thus minimize the sum of the $\chi^2$ and another quantity representing *a priori* belief such as smoothness ($\sum_j \tilde{u}_j^2$) or negentropy ($\sum_j \tilde{u}_j \ln \tilde{u}_j$).

It should be pointed out that the $\chi^2$ fitting is based on the assumption that the counts $[c_i]$ in all pixels are large enough to follow approximately Gaussian distributions. In astronomy, it is rather common that counts in some pixels can be zero or just few. If this is the case, the regularization has to use more accurate statistics (e.g. the Poisson statistics). Such a change brings a great difficulty to the regularizations. In contrast, iterative deconvolution methods do not have this shortcoming since calculations of the Poisson likelihood are after the model estimations.

### 2.2. The Richardson-Lucy Method

The Lucy deconvolution gets an estimate $\tilde{u}_j^{(r)}$ of $u_j$ at iterative step $r(= 1, 2, ...)$ according to

$$\tilde{u}_j^{(r+1)} = \tilde{u}_j^{(r)} \sum_i \frac{c_i}{\tilde{c}_i} p_{ij}, \quad (4)$$

$$\tilde{c}_i = \sum_j \tilde{u}_j^{(r)} p_{ij}. \quad (5)$$

here $\sum_i \tilde{c}_i = \sum_i c_i = C$ (i.e. conservation of total counts); the initial estimate $\tilde{u}_j^{(0)}$ is usually a constant. The Richardson-Lucy method improves the fitting of the estimates step by step and in principle, after an infinite number of iterations, one should derive a convergent signal model $[u_j^\infty]$ that is the mathematical solution of Eq. 1 without noise. In practice, the approach to model $[u_j^\infty]$ is usually very slow, and just because of the noise in $[c_i]$, it is not so useful. We need to give a statistical description of the goodness-of-fit. Up to now, there is not yet a quantitative measure for the probability significance levels of the estimates nor of the statistical acceptability of the solutions. When to stop the iterations is usually judged by one's experience or by *eye impression* in 2-D cases.

In his original paper, Lucy (1974) proposed to use the $\chi^2$ test. This is useful when photon counts are high enough, but may be less sensitive to systematic deviations in the fitting. Recently, Lucy (1994) proposed to use the Kolmogorov-Smirnov (KS) test that measures a least upper bound on the fitting errors. Since the calculations of the $\chi^2$ and KS probabilities are very fast, it is interesting to check at which condition they will become reasonable approximations of the actual significance level that we will derive as follows.

Let us consider the significance levels here. Again, if all $c_i \geq 10$, we can use the $\chi^2$ defined above. But this simplification is not always suitable. In a general case, when type I noise does not exist, a particular configuration (or an imaginary observation) $[d_i]$ with $d_1$ observed in pixel 1, $d_2$ observed in pixel 2 and so on arising from the theoretical distribution $[\tilde{c}_i] \equiv \tilde{c}_1, \tilde{c}_2, ..., \tilde{c}_N$ will follow the multinormal probability function :

$$p([d_i]) = \frac{C!}{C^C} \prod_i^N \frac{\tilde{c}_i^{d_i}}{d_i}, \qquad (6)$$

$$\sum d_i = C. \qquad (7)$$

The mono probability for the real observation $[c_i]$ can be calculated by replacing $[d_i]$ with $[c_i]$ above. The likelihood $M$ of $[d_i]$ can be further defined as

$$M([d_i]; [\tilde{c}_i]) = \ln p([d_i]). \qquad (8)$$

Intuitively, an imaginary observation $[d_i]$ is said to fit to the model $[\tilde{c}_i]$ better than $[c_i]$ if it has the likelihood $M([d_i]) > M([c_i])$; vise versa, if $M([d_i]) < M([c_i])$, it happens more unlikely. So, how frequently better (or worse) imaginary observations appear from a given model $[\tilde{c}_i]$ measures the goodness-of-fit. But such a calculation is impossible to derive by hand because of the complicated forms of Eqs. 6-8.

*2.3. Test for Goodness-of-fit*

We propose the following bootstrap method to access the probability significance level. At each step $r$, we simulate 999 random realizations of $[\tilde{c}_i]$ and derive their associated likelihoods according to the multinormal point process and Eqs. 6-8. Since the total counts are fixed in the Richardson-Lucy algorithm, we use the multinormal process instead of the Poisson process. If this constraint is relaxed, the Richardson-Lucy routine may become uncontrolled because the direction of the Poisson error is unknown. In practice, the total number of counts is always large so the difference between the two processes is negligible.

A random sample is created in the following ways : 1) generate a random number $r$ between 0 and 1; 2) by comparing it to the cumulative distribution function of $[\tilde{c}_i]$, we add one count to grid $j$ where $\sum_1^{j-1} \tilde{c}_i < r \leq \sum_1^j \tilde{c}_i$; and 3) repeat the above procedure $C$ times. As one could expected, there are several numerical routines that can compute step 2 at the speed $\log N$ per count.

Plus the observed $M([c_i])$, there are totally 1000 likelihoods. The idea here is to sort them in an ascending order, $M(1), M(2), ..., M(K), ..., M(1000)$, with order $K$ denoting the real data. The null hypothesis is that $[c_i]$ comes identically from a random realization of the model $[\tilde{c}_i]$ against the alternative that this is incorrect. If $K = 1$, the real data appear most *unlikely*, so the hypothesis should be rejected at the 99.9% confidence level. In general, the chance or the probability significance level for getting the real data out of the model is

$$prob = K/1000. \qquad (9)$$

Because the Richardson-Lucy algorithm improves the likelihood step by step, $K$ will increase from 0 to 1000. One sees that models of both small $K$ ($\simeq 1$) and high $K$ ($\simeq 1000$) should be rejected : a small *prob* means a poor fitting while a large *prob* (corresponding to $\chi^2 \simeq 0$) means the fitting is too good because one has collected all noise as signal. All the simulation models and the real ROSAT X-ray spectral data we have investigated show a behavior as in Fig. 5, say, after number of iterations the probability reaches a certain value and stays almost at it even if many more iterations are carried out. This level depends on the point spread function as well as the structure of the underlying signals. Whether it can be used for determination will depend on one's *a priori* entropy knowledge which we discuss in section 2.4. But at least, a reasonable model should reach a minimal value for this level. From our experience, we suggest $prob = 20\%$ for this minimal value.

Even with $K = 1000$, bootstrap sampling errors may be still visible. If they become too serious, one should try even more random samples. In principle, the exact significant level will be approached at $K \to \infty$. However, for practical use, the number of $K = 1000$ is already enough.

We should stress that the test given above is a global measure of the goodness-of-fit. As soon as a global significance level, $g$, is derived, *every* part of the model will fit to the data locally at a level larger than or equal to $g$, a consequence of Eq. 6. This condition is important since the Richardson-Lucy and related iterative methods converge locally.

*2.4. Entropy*

While the Richardson-Lucy algorithm finds model estimates of significance levels from zero to 100%, the opposite is not true — not all models in the range can be scanned by the algorithm. Because of noise, imperfect $p_{ij}$ and oversampling, one observed data set does not constrain a unique deconvolution model at a given significant level. Iterative algorithms, e.g. the Richardson-Lucy method, give a practical way to search for a particular, smaller set of all possibilities. However, the models scanned by the proposed algorithms are always relatively smoother or of higher entropy in the signal space and thus are more physically reasonable than the rest. In the Richardson-Lucy method, such an important property is guaranteed by starting from the smooth maximum-entropy initial condition.

Adding more iterations in the Richardson-Lucy algorithm improves the significance but degenerates the entropy of the models defined as :

$$E = -\sum_j^M (\tilde{u}_j \ln \tilde{u}_j + \ln \tilde{u}_j), \qquad (10)$$

in the absence of any other *a priori* knowledge (Narayan & Nityananda 1986; Skilling 1989; Press etal. 1992). Equiv-

alently, the *a priori* knowledge is just that the signal is as smooth as possible. The entropy that characterizes smoothness or variance of the unknown signals can be used as an implemental quantity in the deconvolution. If more steps improve the goodness-of-fit very slowly, but loose too much entropies, one probably should stop the iteration at a lower significance level.

We should point out that iterative algorithms that maximize both the mono probability $p$ (Eq. 6) and the entropy (Eq. 10) do exist (Frieden 1972; Hunt 1977 and Meinel 1986), but the solutions *do not guarantee an acceptable significance level in the observation space*. For instance, for the ROSAT spectrum of G0905-098 below, this maximum solution is derived after the first iterative step in the Richardson-Lucy algorithm, but the model has only the significance level 3%. Therefore, maximum-entropy algorithms should not be preferred to the knowledge-free Richardson-Lucy method if the *a priori* knowledge is too far from the fact (as the smooth *a priori* knowledge, Eq. 10, sometimes does). A good way may be to start from the Richardson-Lucy algorithm and after a reasonable significance level is achieved, one updates the *a priori* knowledge and uses a maximum-entropy method. Our studies for 1-D ROSAT spectra show however only a slight difference between this and the simplest Richardson-Lucy method.

### 2.5. CPU time consumption

The CPU time for the bootstrap is dominated by simulating random samples which is linearly dependent on $999 * C * \log N$. On our IBM-RISCS workstation (128M RAM, multiusers), it takes a few minutes to get one significance level at one iterative step for $C = 1000$ and $N = 256$. For 1-D spectra, this is not a problem because the convergence is usually obtained at the first few steps. In 2-D images, a good model may require few tens of iterations and then the calculations are performed at selected steps. Anyway, this is well within present computation power.

## 3. Applications

### 3.1. A simple model

This is a very simple, but still a non-trivial, example :

$$c_1 = 0.75 u_1 + 0.25 u_2 + \text{Poisson noise} (c_1), \tag{11}$$

$$c_2 = 0.25 u_1 + 0.75 u_2 + \text{Poisson noise} (c_2), \tag{12}$$

here the point spread function is $p_{1,1} = 0.75, p_{1,2} = 0.25, p_{2,1} = 0.25, p_{2,2} = 0.75$. The real signal is assumed to be $(u_1, u_2) = (20, 40)$, so 25 counts and 35 counts are predicted in channel 1 and 2 respectively. But due to Poisson statistics, we observed $c_1 = 24, c_2 = 38$.

The Richardson-Lucy deconvolution starts from $u_1^0 = u_2^0 = 31$. Fig. 1 illustrates from bottom-up the changes of $\tilde{u}_1$ (circles + line), $\tilde{c}_1$ (triangles + line), $\tilde{c}_2$ (triangles) and $\tilde{u}_2$ (circles). The signal estimate converges to $(\tilde{u}_1^\infty, \tilde{u}_2^\infty) = (17, 45)$, the solution of the linear equations :

$$c_1 = 0.75 \tilde{u}_1^\infty + 0.25 \tilde{u}_2^\infty, \tag{13}$$

$$c_2 = 0.25 \tilde{u}_1^\infty + 0.75 \tilde{u}_2^\infty. \tag{14}$$

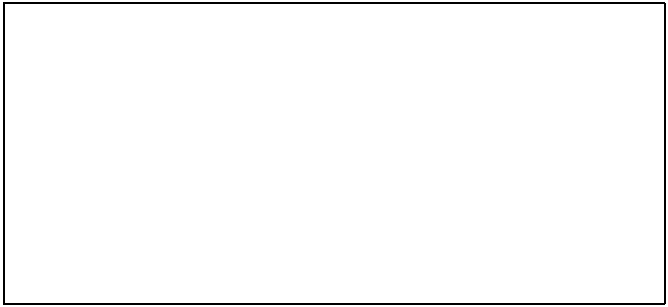

**Fig. 1.** the deconvolved results in the simple model. The lined circles are for $\tilde{u}_1$, the lined triangles for $\tilde{c}_1$, the circles for $\tilde{u}_2$ and the triangles for $\tilde{c}_2$.

After about 15 steps, the estimate (17.4,44.6) is already close to the mathematical convergence. At step 40, it is $\tilde{u}_1 = 17.001$ and $\tilde{u}_2 = 44.999$. The real signal (20,40) cannot be restored because of the noise.

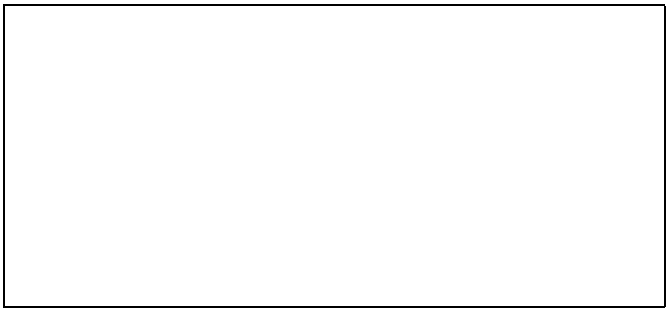

**Fig. 2.** simulated significance levels via iterative steps. Note that the probability should be a monotonous increasing function in principle; the actual fluctuations (e.g. at step 15) are due to sampling errors.

We have used the bootstrap simulations to obtain *prob* in Eq. 9. Fig. 2 shows how the significance level *prob* increases with $r$ and Fig. 3 how the entropy $E$ decrease with *prob* (the so called trade-off curve, see Press etal. 1992). There are still small sampling errors due to the finite bootstrap in the figures but these are easily checked out. If adopting $prob = 0.6$, we can stop at step 4 when $\tilde{u}_1 = 21.6, \tilde{u}_2 = 40.4$. This model is even closer to the real signal (20,40) than the convergent solution (17,45). The signal range corresponding to the significance levels $prob = 0.3 \rightarrow 0.8$ are $u_1 = 24.9 \simeq 19.7$ and $u_2 = 37.1 \simeq 42.3$.

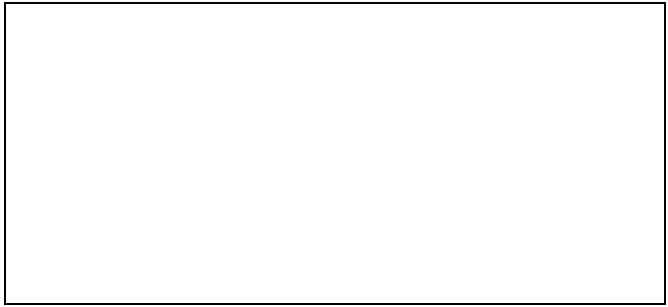

**Fig. 3.** illustrations of how the entropies are changed with the probabilities. These are so called trade-off curves, see the text for more.

We note that the *range* of the modeled signals depends on the initial condition. For instance, if $u_1^0 = 5, u_2^0 = 57$ are used, the range $u_1 = 24.9 \sim 19.7$ and $u_2 = 37.1 \sim 42.3$ derived above can never be recovered. The important point here is that models coming from this irregular initial condition have entropies less than those corresponding to the smooth initial condition. One should always use the smooth condition. It is not a mathematical reason, but simply scientists' belief, that physical processes in the Nature should be as smooth as possible, or have entropies as large as possible.

### 3.2. ROSAT 1-D spectra

We apply the above bootstrap method to ROSAT X-ray spectra. The ROSAT telescope receives soft X-rays between 0.07 Kev and 3 Kev with an effective area depending on photon energy, and records them as *photon events* in 256 PSPC detector channels from 0.08 Kev to 2.5 Kev. The detector response matrix (the PSF) is well calibrated for 728 incident energies and 249 output channels. But to be not too oversampling, we usually rebin the 728 input energies into 249. For a source of counts less than 500, one needs to rebin them further in order to get acceptable estimates quickly.

The X-ray spectrum comes from the radio-jet AGN galaxy G0905-098 ($m_v = 15.5$), found to be X-ray bright by EINSTEIN and re-observed by ROSAT. It is situated at the boundary of the Abell cluster A794 which itself has bright intracluster X-ray emission. Because of the particular radio configuration and the particular site, it was an ideal example to study the physical interaction between the intracluster medium and the radio jet. The raw X-ray spectral histogram is shown in Fig. 4.

We start from a smooth initial estimate normalized to have the same total counts 2178 as observed. Fig. 5 shows how the significance level increases. The biggest improvement of the model happens at step 2 where the probability is rapidly increased from 3% to 25%. After step 2, the fitting of $[\tilde{c}_i]$ to the data $[c_i]$ is improved very slowly although in the signal space $[\tilde{d}_j]$ becomes more and more irregular.

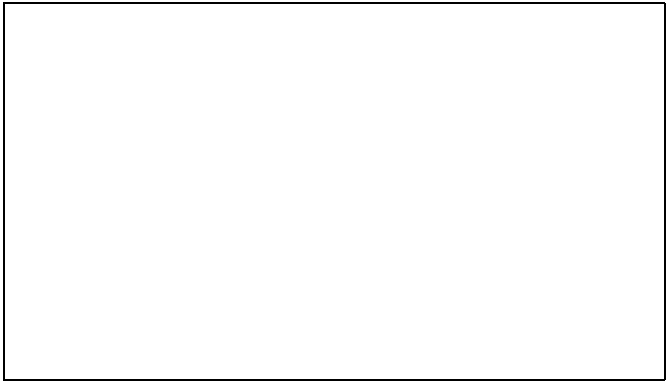

**Fig. 4.** the raw counts of the G0905-098 ROSAT spectrum are plotted as the histogram. The curve is the objective-feasibility filtering from the Lucy deconvolution at step 2.

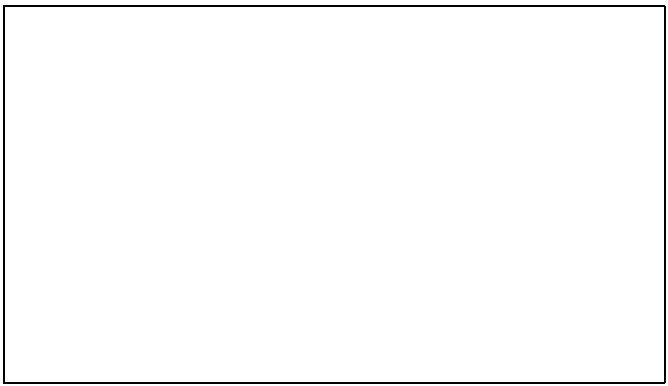

**Fig. 5.** similar to Fig. 2, but here for the ROSAT 1-D spectrum. Note that the improvement of the fitting after step 2 is very slow.

Apparently, the best guess for the model should be at step 2 because the model is much smoother while still at the significant level 25%. The light curve in Fig. 4 shows the object-feasibility filtering $[\tilde{c}_i]$ and the histogram in Fig 6 the corresponding deconvolution model spectrum $[\tilde{d}_j]$. It can be explained simply by a power law AGN spectrum plus the galactic absorption (seen as the turn-over at 0.25 Kev). We note that some details do appear after about 40 step with the significance level about 30%, but since this level is not very different from that at step 2, the details are not very reliable and should be checked independently.

In Fig. 7, the circles are for the likelihoods and the triangles for the entropies of Eq. 10 (both normalized to be 0 at step 1, so the initial values at step 0 are outside the figure). The curve gives the sum. It is important to note that the maximum of the curve is reached at step 1 that corresponds to the significant level (3%) — too low to be acceptable. Thus, the solution that maximizes both the likelihoods and the entropy fails to fit to the data ! We used the entropy-based iterative algorithm in Meinel (1986) and have confirmed this result. In order to do maximum-entropy deconvolutions, *a priori* knowledge of the Richardson-Lucy estimate at step 2 has to be updated. The follow-up deconvolution results are similar to but slightly more stable than those derived from the simple Richardson-Lucy method. Of course, such a continuation is needed only if it can significantly improve the significance level (not in this example).

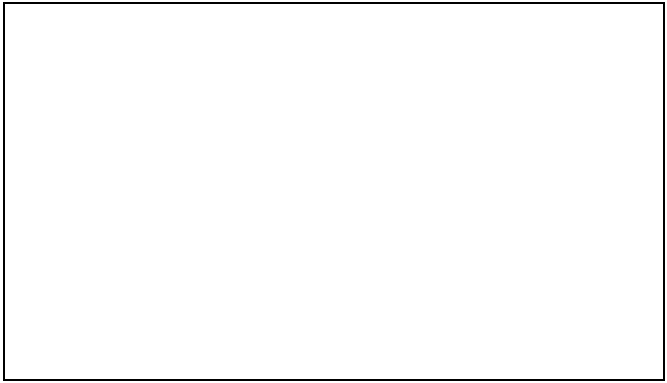

**Fig. 6.** the deconvolved X-ray spectrum at step 2.

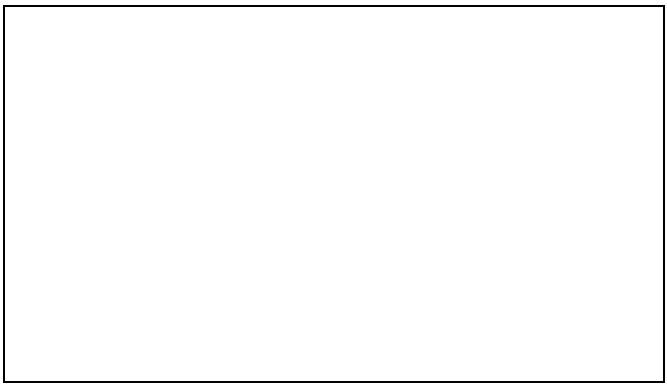

**Fig. 7.** illustrations of how different likelihoods (the circles representing for log mono probabilities and triangles for entropies) changes with iterative steps. Maximum-entropy methods are usually designed to find the maximum of the sum of the two likelihoods which we plotted as the line curve. But in this example the maximum is reached at step 1 which has a very poor significance level, say, 3%.

### 4. Discussions and Summary

We have proposed a bootstrap method to calculate the probability significance levels of deconvolution models in the Richardson-Lucy algorithm. This is a global measure for the method that converges locally. The bootstrap usually takes a considerable computer CPU time but is still tolerable by present powerful workstations. With this method, one can calculate at each Richardson-Lucy iterative step how good the model is, and judge when the iterations can be stopped.

Though more iterations give better fitting in the sense that the significance levels are improved, such a progress is usually very slow after first tens or even few steps, while it may loose too much entropy or smoothness. In principle, the convergent model of the $\simeq 100\%$ significance level can be derived by an infinite number of iterations. But in practice, this is neither possible nor useful because such a model would have taken noise completely as real signals. A reasonable model can be chosen at the significance level 20% to 80%.

We have also discussed how to incorporate the entropy in the deconvolution. In some cases, if one maximizes the sum of the entropy described by Eq. 10 and the model likelihood (Eq. 9), the solution does not guarantee an acceptable significance level (the data are fitted too poorly). Therefore, it is recommended that *a priori* knowledge of the entropy derived from a few Richardson-Lucy iterations should be updated accordingly, if this is necessary in improving the significance level quickly.

The above convolutions are illustrated in our two 1-D spectral examples. Detailed applications to 2-D data will be published elsewhere.

*Acknowledgements.* We are grateful to L. Lucy for stimulating discussions on iterative algorithms, statistical significance levels and maximum entropy. G. Hasinger and H. Zimmermann are acknowledged for their very useful suggestions about the bootstrap simulation and ROSAT spectral measurement.

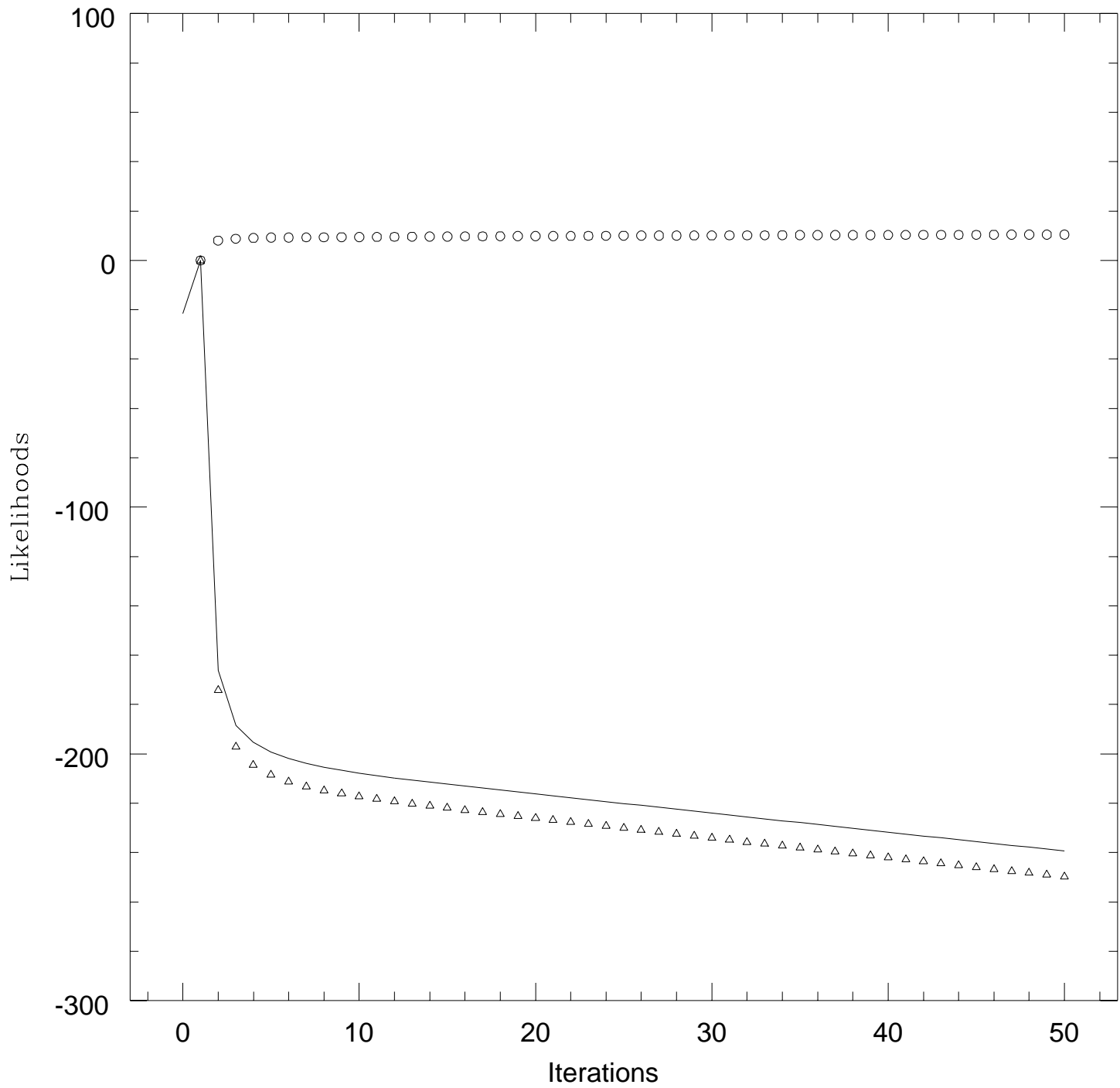

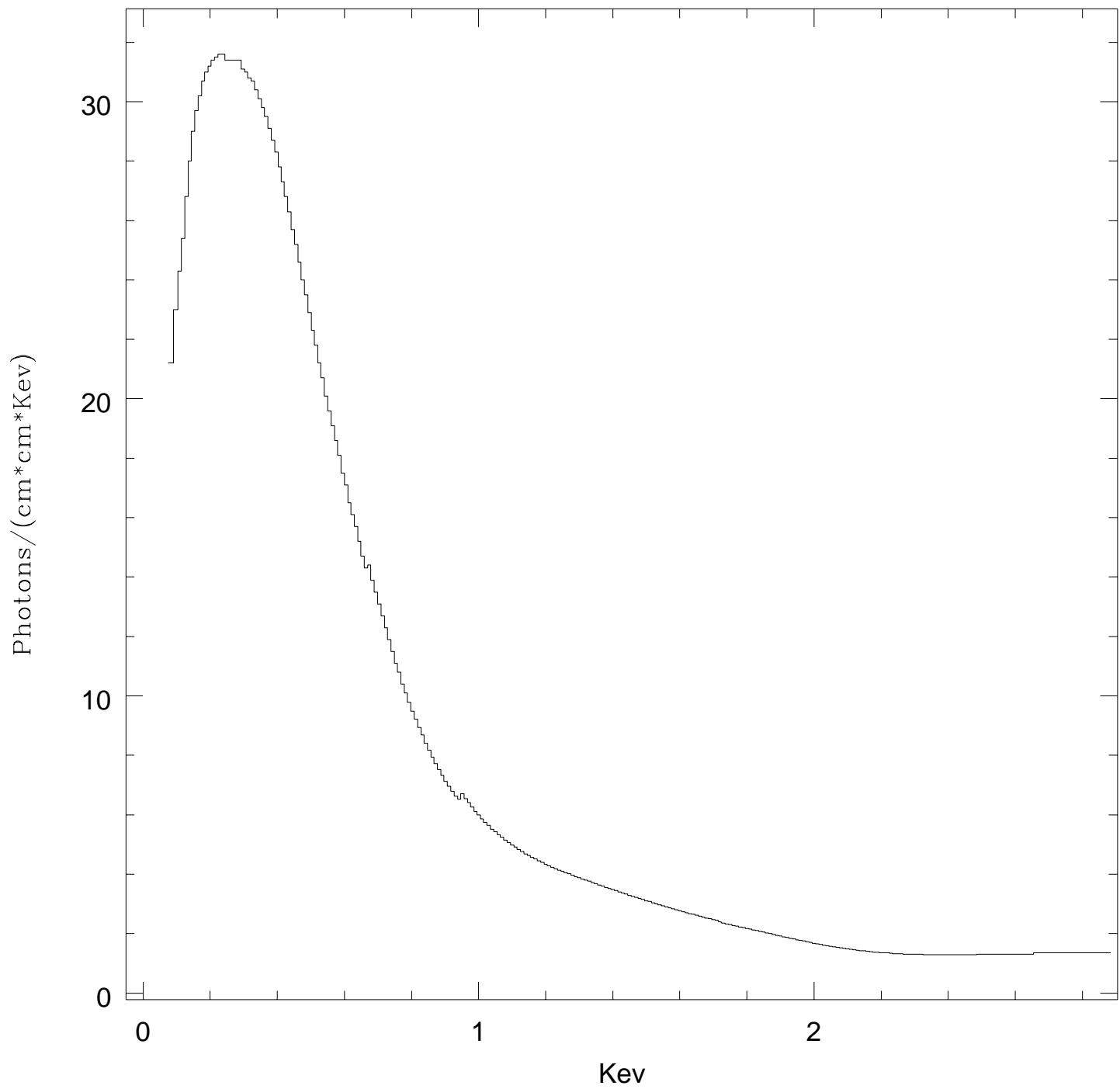

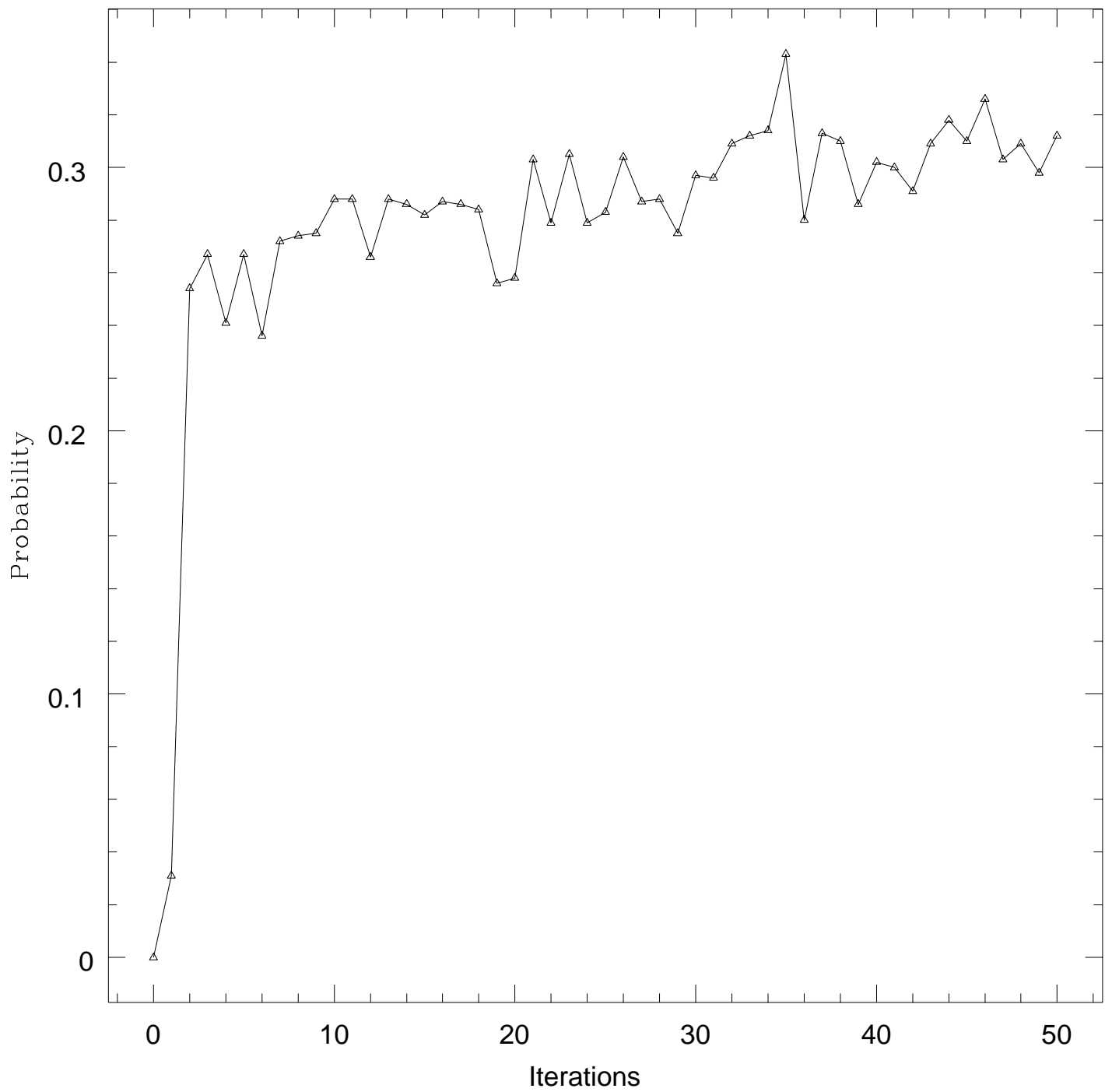

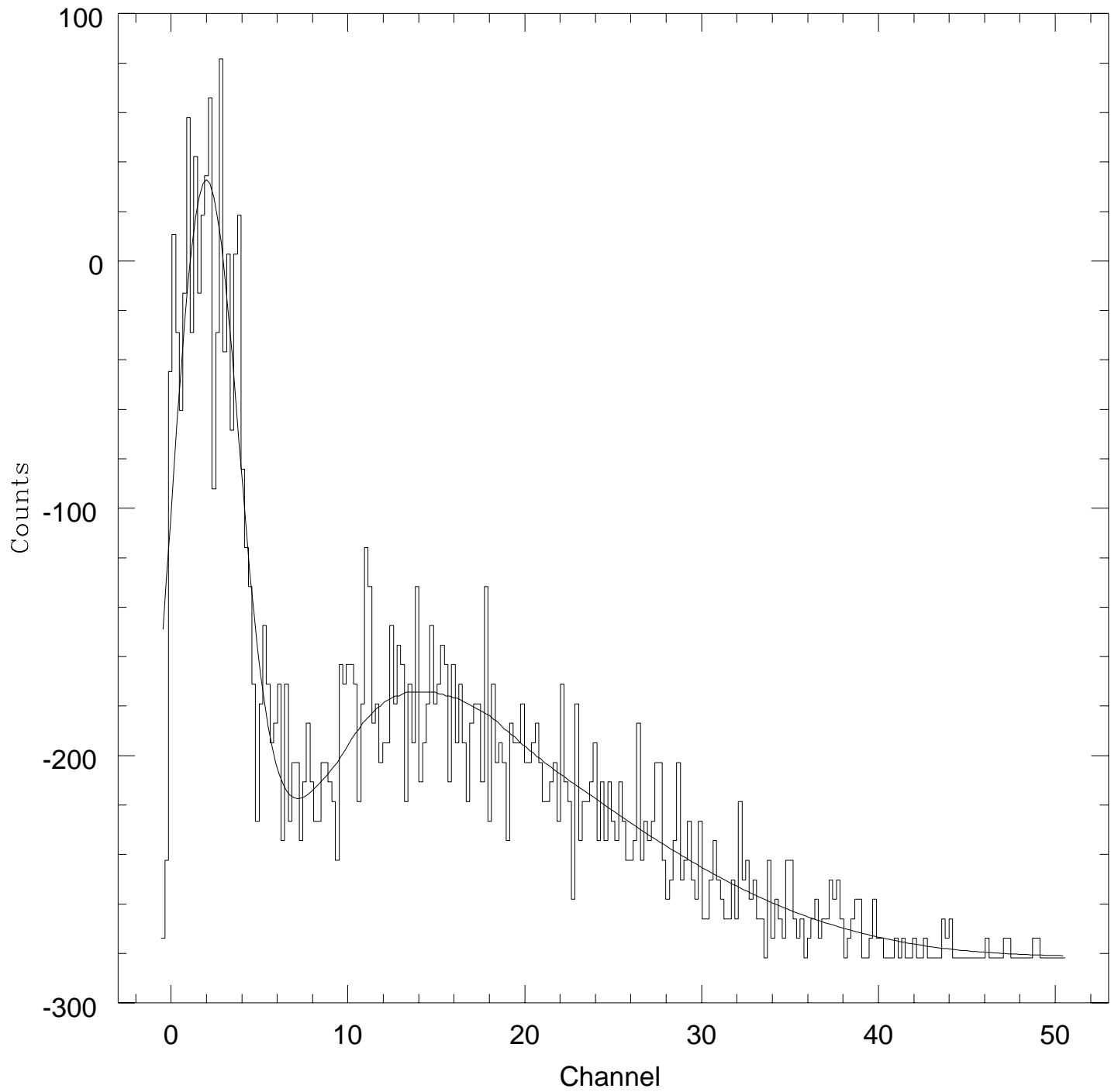

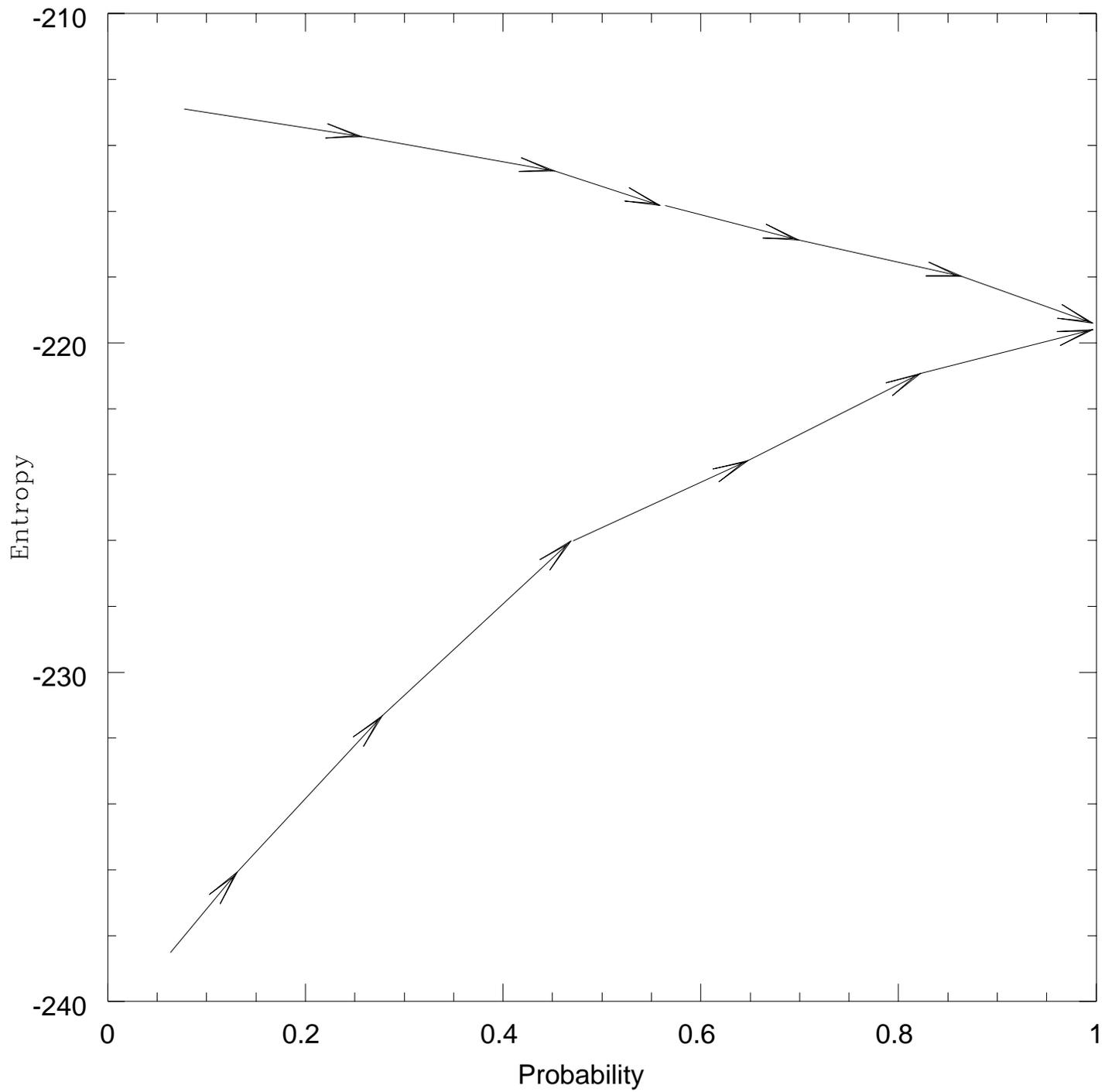

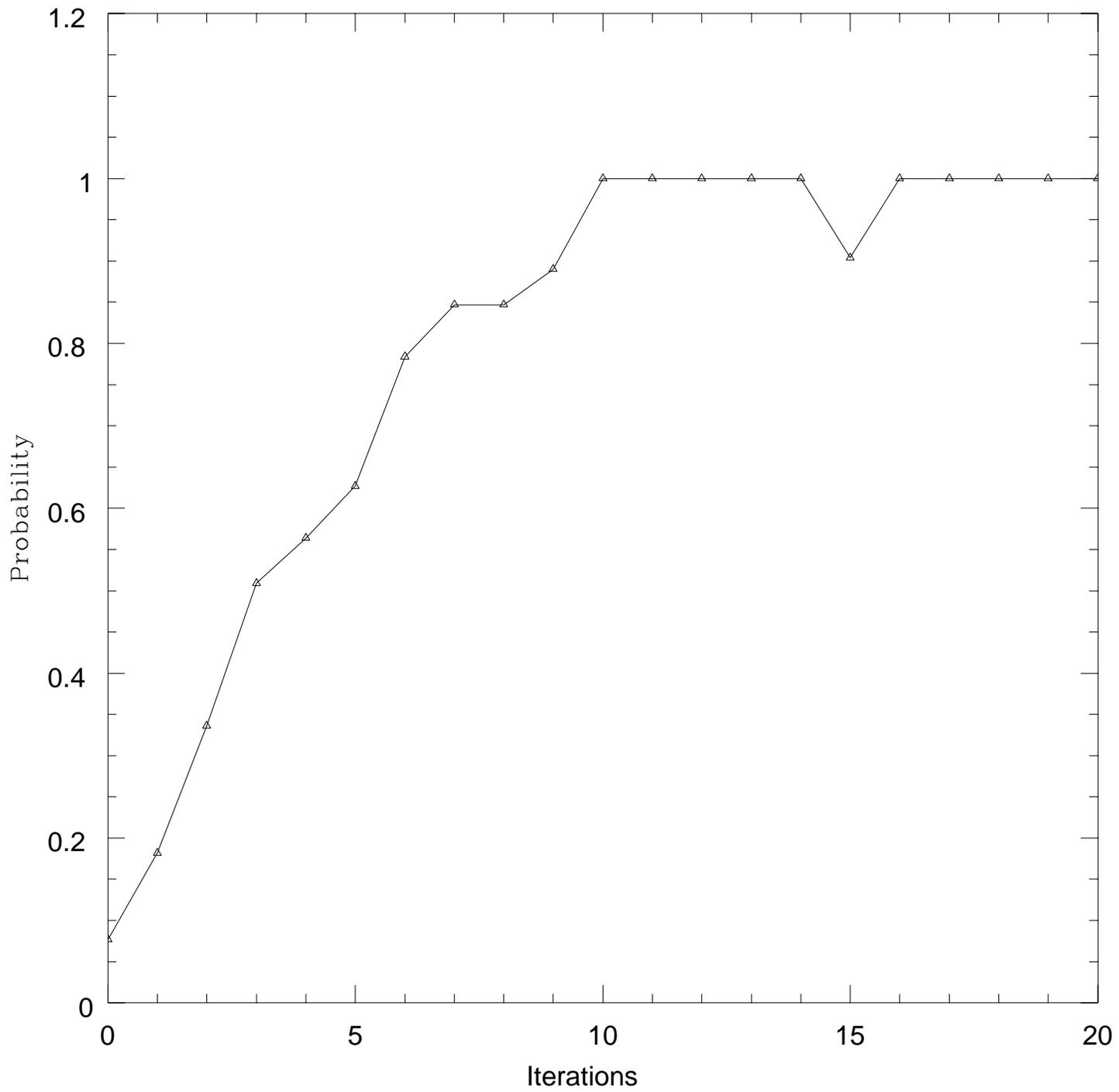

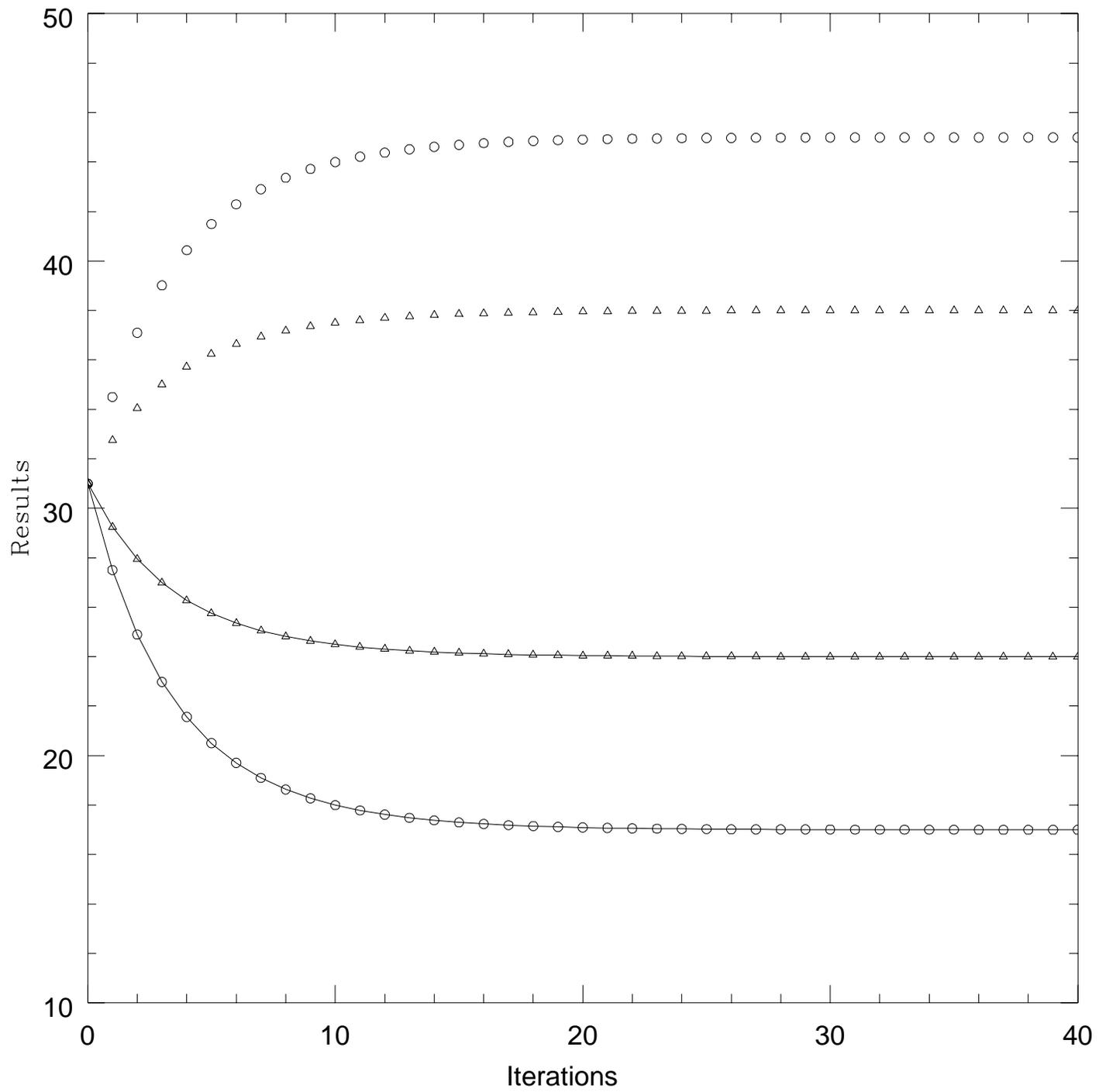